# A COMPARATIVE ANALYSIS OF THE CYBER SECURITY STRATEGY OF BANGLADESH

Kaushik Sarker, Hasibur Rahman, Khandaker Farzana Rahman, Md. Shohel Arman, Saikat Biswas, Touhid Bhuiyan

Dept. of Software Engineering, Daffodil International University, Dhaka, Bangladesh

## ABSTRACT

*Technology is an endless evolving expression in modern era, which increased security concerns and pushed us to create cyber environment. A National Cyber Security Strategy (NCSS) of a country reflects the state of that country's cyber strength which represents the aim and vision of the cyber security of a country. Formerly, researchers have worked on NCSS by comparing NCSS between different nations for international collaboration and harmonization and some researchers worked on policy framework for their respective governments. However very insignificant attempts had been made to assess the strategic strength of NCSS of Bangladesh by performing cross comparisons on NCSS of different Nations. Therefore, the motive of this research is to evaluate the robustness of the existing cyber security strategy of Bangladesh in comparison with some of the most technologically advanced countries in Asian continent and others like USA, Japan, Singapore, Malaysia and India in order to keep the NCSS of Bangladesh up-to-date.*

## KEYWORDS

## 1. INTRODUCTION

Internet has established itself as a fundamental and essential necessity for the life of the people and for their socio economical activities. Though it is making things easier for people it's also bringing new emerging risks. Cyber-attacks have never stopped and never will, instead it's increasing exponentially. Therefore, every country requires a stable, reliable and resilient ICT infrastructure [1]. A weak ICT infrastructure can be of high risk. Utilizing ICT, any interested party expert in cyber intelligence can exploit classified information about Government and Industries with advanced technologies. The government of Bangladesh planned to achieve 7.2% GDP growth in the year 2016-2017 [2]. To touch the desired target ICT sector has an important role to play. Therefore, the existing cyber security strategy has a key role in the social, economic and national development process by reducing cyber threats, providing economic security, strengthening national resiliency, political imperativeness, lawful mandate, protecting state secrets, strengthening diplomacy increasing country image, etc. As the NCSS of Bangladesh was declared in 2014, it's been high time to check the strategy if it is still up-to the task to deal with the emerging national and international threats. Researchers have worked on different levels of security strategies to improve their respective areas. Developed countries have worked on their security frameworks, policies both in international and national perspective. As Bangladesh followed the ITU framework for the creation of NCSS, it can be considered that it has complied with the international policies. However, very insignificant national level interest to check the efficiency of the NCSS has been noticed like framework analysis, policy analysis, implementation state, current progress etc. Therefore, this research is an attempt to find the current state of the NCSS of Bangladesh by comparing the strategy with some countries of different technological progress level like USA, India, Japan, Malaysia, Singapore and find some insights through the comparison to improve the strategy.

The organization of chapters is as followings. After the introduction in the first chapter, the second chapter describes the background of this study, and literature review in the next chapter. After that, methodology of this research has been specified with the mention of comparison criteria which have identified as the effective comparison criteria to compare a cyber-security strategies. In the next chapter findings and analysis is done which shows the cross section analysis of the strategies of specific countries based on the comparison criteria. Then results from the findings have been discussed with some recommendations and finally the conclusion has been drawn with the notion of future work.

## 2. BACKGROUND

In this decade the number of Internet user has increased exponentially. It's happening to the whole globe. From underdeveloped to developing and developed country, every single country is now bested with the blessing and curse of the technology. It's increasing and along with this, cyber-attacks are also getting more advanced and complex. To make things worse, unlike the federal crimes, it's not restricted to specific country, it's beyond border. Thus, it's more likely to remain undetected if proper infrastructure is not there to fight it. Bangladesh government's vision is to transform helself into a digital one. That suggests more and more involvement with technology. Previously there have been a good number of cyber- attacks on her cyber space. In 2012, 26 government organizations were hacked. In 2013 one government bank (Sonali Bank) was under attack and the hackers were able to cart away US$250,000 without making any fuss [3]. In 2015 the web sites of Bangladesh police and the Rapid Action Battalion (An Elite force) were hacked February 6 2015 [4]. These are a few as examples. However, the threat of cyber intelligence was mainly triggered by the 2016 Central Bank heist. It was one of the world's biggest bank heist of cybercrime history. They were able to steal $101 million from the Bangladesh bank account at the Federal Reserve Bank of New York through the SWIFT network [5]. Later it was found that Dridex (A banking Trojan) was used in this heist [6].

The Cyber security strategy of Bangladesh was launched on 2014. And its mission is to protect the cyber world against security threats, risks and challenges to national security. It is in accordance with the IMPACT (International Multilateral Pact Against Cyber Threats). The motivation is to create a vision for keeping the country secure in government sectors, private sectors, citizens and international cyberspace. However, after a heist of such a large scale Bangladesh can easily assume the security condition of her cyberspace. It's high time to take a look at the cyber security strategy of Bangladesh that is currently in place. The approach that has been taken to achieve the goal is pretty straight forward. Different countries have taken different measures to make their NCSS better. As the motive is to know the present situation of the NCSS of Bangladesh, it seems a good plan to compare it with some already strategically strong countries and some less strong ones to find Bangladesh's position in that chart.

## 3. LITERATURE REVIEW

Various research contributions towards cyber security had been made formerly. However the comparative research that this paper is focused on is believed to be able to fulfil a research gap. To state the motives behind developing cyber security strategy three individuals Azmi, Tibben and Win orchestrated a literature review of National Cyber Security Strategy on global context. The review [7] used qualitative comparison approach between 54 countries and the findings shows key reasons behind creating cyber security strategy. It implies the importance for creating cyber security strategy. The main drawback was, the conclusions they reached were based on the study of existing NCSS themes. Therefore the reliability of the themes could not be ensured. Another Case study performed by the National Cyber Bureau of Israel on Cyber security policy model proposed a framework for the creation of cyber security. The case study [8] was a comparative research between the NCSS of countries that are considered world pioneers in technology. The study was conducted by Deniel Benoliel. The study found some best practices that were followed by all the countries involved in the cross comparison of the study. However the limitation of the study is every country has some different agenda, therefore they certainly will have some differences while creating their strategies due to distinct national and international interests. The framework is flexible which can be considered as a positive aspect. Similar works in the field of cyber security strategies have been orchestrated by different authors. However, very insignificant attempts had been made formerly to evaluate the NCSS of Bangladesh in context with the other countries. Therefore this research is going to cover the research gap found in this context. As far as the national cyber security strategy of Bangladesh is concerned the strategy outlines a framework for organizing and prioritizing efforts to manage risks of Bangladesh's cyberspace which directly follows the GCA (International Telecommunication Union's Global Cyber Security Agenda). In line with the GCA, this strategy implies the prioritization of the strategic areas. These areas are – National Cybercrime legislation, reducing vulnerabilities in software products, raising awareness and International cooperation.

## 3.1 Priority 1: Legal Measures

The first priority is focused on the strategies that are involved with the development of cybercrime legislation that is globally applicable. The priority deals with the enactment of laws to prosecute and deter cybercrime, establishment and modernization of laws which will depend on local needs and national condition.

### 3.1.1 Cybercrime Legislation

In this action the strategy implies that cybercrime legislation should comply with the global conventions in order to be interoperable and applicable globally. And the strategy explicitly mentioned ITU as the matrix for the alignment of the policy with global conventions. After building up the law, the law should be evaluated by all ministries and legislative committees so that there's no usable information missed. In accordance with the strategy, the Cyber Security Act 2015 has been proposed to fortify the cybercrime legislation [9].

### 3.1.2 Government Legal Authority

Creation of law is the first step towards strengthening our cyber security. Next is the implementation of that law. To make it happen, government needs proper legal authority. This involves creating cyber security organizations (e.g. - National Cyber Security Council).

## 3.2 Technical and Procedural Measures

This priority involves the rules and regulations for creating organizational structures. Any organization or company should follow specific regulations set by government. Bangladesh has set the regulation on both the CIRT [10] and this strategy suggested the below actions as the technical and procedural measures.

### 3.2.1 National Cyber Security Framework

This action is aimed towards the creation of mandatory security standards and guidance on issues such as compliance, assurance and risk management. This is the framework that every tech related company must adopt. In the first International Cyber Security conference in Bangladesh that took place on March 2017, government has taken some declaration that involves guideline or instruction for organizations and stakeholders [11].

**3.2.1.1 Secure Government Infrastructure**

This section deals with creating awareness of risks and preventive measures. This is targeted towards government departments and agencies. Building a strong security infrastructure involves awareness of mass people.  Government has established National data centre and taken some initiatives in hand to create some high-tech parks. Clearly the numbers of critical infrastructures are increasing. Therefore the need of the protecting of those too is increasing.

## 3.3 Organizational Structures

This section involves the strategy for building the organizational structures to help prevent and respond to attack against critical infrastructure.

### 3.3.1 Government's Cyber Security Role

The strategy suggests that the government should take the lead in securing the cyberspace. Securing a country's cyberspace involves not only the government but also the collaboration of private sectors. The strategy mentions about appointing an aid as coordinator, who can act as the head of the National cyber security.

### 3.3.2 National Cyber Security Council

To coordinate this national effort the government of Bangladesh has taken steps to create a multi-council body, which is known as National Cyber Security Council. This council involves different roles such as developing national plans [12], incident response capacity, strategic advice [13] acting as national authority for information assurance, collaborating with international schemes [14] and perform R&D.

### 3.3.3 National Incident Management Capacity

This section of the strategy deals about the incident management capacity. Prevention is always better than cure. However, there's no way anyone can be sure that any large scale prevention method can defend or prevent attacks. This strategy portion has discussed about the damage control policy after any possible incident takes place. For any kind of cyber related incident response, Bangladesh government has established an Incident response team named BD-CIRT [15].

### 3.3.4 Public Private Partnership

It's an important fact that, not only government can handle the threats that they are against. As cyber space is an open platform, it's more dependent on the private sectors. Therefore private sectors have a great deal of influence on Bangladesh's cyber space. This section of the strategy dictates the terms on private- public relationship. Bangladesh government has issued policy on different aspects of cyber space both on business level and security level for private organizations. It is mandated that the private sectors must obey the policy been imposed. This policy involves the collaboration of work and resources, exchange of information and regulation for stakeholders between government and private sectors [16].

### 3.3.5 Cyber Security Skills and Trainings

This section covers the security related trainings and skills framework. Bangladesh government has taken multiple steps to increase the number of IT professionals within 2021. They have taken projects on training and creating skilled IT professionals. Though here the need for Cyber security skills and trainings are mentioned, no particular project to create skilful cyber warriors has yet been started.

### 3.3.6 National Culture of Cyber Security

Cyber security awareness can reduce the impact of cyber-attacks at great extent. The way internet is growing, awareness about its crimes has not grown in the same manner. This section of the policy discusses about necessary steps to take to create awareness among mass people and government employees.

## 4. METHODOLOGY

### 4.1 Research Type

This research requires gathering relevant information from the specified documents related to the intended topic. Therefore, qualitative research will be followed in this study in order to analyse and come to an understanding about the cyber security strategy of Bangladesh.

### 4.2 Method to Carry Out Research

The research method that will be used is analysing other country's cyber security strategy in a cross comparison with the cyber security strategy of Bangladesh that will help to understand the strength of the cyber security policy of Bangladesh. A brief review of the strategy of Bangladesh will be done prior to the cross comparison between other countries. Five other countries have been chosen as the comparison vector. Cyber security strategies of USA, Japan, Malaysia, Singapore and India will be studied and analysed. Those analysed data will be categorized into some criteria that are already been suggested as standards for a cyber-security strategy framework by the Israeli National Bureau of Cyber security. Those categorized information will then be compared side by side with the cyber security strategy of Bangladesh.

### 4.3 The Comparison Criteria

Strategic differences cannot be avoided. However, it has been found that whichever level of threat a country is facing, if it has to survive the cyber space in a secure manner, these few categories are known to be best practised and to be included in the cyber security strategy of a particular country as these have been already been a part of the cyber security strategy of some of the highly technologically advanced countries like the United Stated, United Kingdom, Canada, Netherlands, Japan etc. The practices and declared policies by above mentioned countries suggest the following list of criteria as best practices [17].

### 4.3.1 Promote Cyber security R&D

Such practices will cultivate dynamic research communities that are able to take next generation challenges and it will bring an opportunity for the industries to expand the overseas markets.

### 4.3.2 Promote Cyber security Education

Such practice helps nations gain the required resources and skills to build a resilient infrastructure and technology to achieve desired level of security in cyberspace.

### 4.3.3 Ensuring on going Risk Assessment

For justifying, certifying and to strengthen the cyber security strategy and competence of information security and the policy, this practice should be exercised periodically.

### 4.3.4 Promote Counter Cybercrime Policy

As the international nature of cybercrime goes, to increase the capabilities of the law enforcements and the legislation both in national and international level, this practice helps.

### 4.3.5 Promote Cyber security in International Law

Because of the international nature of cybercrime, it's diversity in language, culture and ideas are noticeable. For the same reason international cyber laws are to be applied. To collaborate to the international laws, this practice is necessary.

### 4.3.6 Forms of Regulation and Institutional Aspects

There must be some technological standards and regulations to be followed to ensure a safe cyber space. This practice is about the regulations to be maintained by institutions both in public and private sectors.

### 4.3.7 Balancing Cyber security with Civil Liberties

Sometimes too much security and regulation scare away the potential users. Therefore this practice is about a less complex cyberspace with proper security.

### 4.3.8 Types of Cooperation

Without cooperation there will be no safe cyberspace. There are four different types of cooperation, namely, Public Private Platform, Inter-Governmental Cooperation, Regional Cooperation, and Intra-Governmental Cooperation.

## 5. DATA FINDINGS FROM THE CROSS SECTION ANALYSIS

This section represents the cross section comparison of strategies. Each table shows the strategy in specific category for all the mentioned countries side by side.

Table. 1. Comparison in Promoting Cyber Security R & D

| Country | 1. Promote Cyber Security R & D |
|---|---|
| USA | (1)To accelerate innovative cyber research and development to build cyber capabilities. (2)It will focus its basic and applied research agenda on developing cyber capabilities to expand the capacity of the CMF and the broader DoD [19] cyber workforce [20]. |
| Japan | (1)For the purposes of Japan maintaining and improving its own leading research and development, the research and development and practical testing of technologies aimed at improving the cyber-attack detection and advanced analysis functions at research institutions and relevant organizations shall be accelerated [24]. (2) To improve the cyber-attack detecting and advanced analysis functions at research institutions, Japan will take some steps to improve its R & D and practical testing of technologies. (3) To promote R&D for cyber security through NICT, National Institute of Information and Communication, MIC operates research infrastructure [25]. |
| Singapore | (1)Singapore's cyber security R & D journey has already started with the aim of translating R & D capability in Singapore into operational strength. (2)Singapore will continue to establish world class R & D facilities in specialized research areas to attract top researchers and international collaborator, and will promote the shared use of such facilities. (3)The government will initiate a cyber-security consortium with S$1.5 million in funding over three years from 2016. |
| | (4)Support research into both technological and human science aspects of cyber security through the S$ 190 million National Cyber security (R & D) Program [26]. |
| India | (1)The R&D sector in India is all set to witness some robust growth in the coming years. According to a study by management consulting firm Zinnia, engineering R&D market in India is estimated to grow at a CAGR of 14 per cent to reach US$ 42 billion by 2020 [21]. (2) Indian IT industry is expected to add to the development of the R&D sector [22]. (3) To undertake R & D programs for addressing all aspects of development aimed at short term, medium term and long term goals including development of trustworthy systems, their testing, deployment and maintenance throughout the life cycle. (4) To make easy transition, pervasion and commercialization or the outputs of R & D into commercial products and services for use public and private sectors [23]. |
| Malaysia | (1)Promote the development and commercialization of intellectual properties, technologies and innovations through focused R & D. (2)Enlarge and strengthen the cyber security research community. (3)Formalize the coordination and cyber security R & D activities [27]. |
| Bangladesh | (1)Prioritize national cyber security R & D activities. (2)Perform and fund R & D with other agencies to create a new generation of secure cyber technologies [18]. |

Table. 2. Comparison in Promoting Cyber security Education

| Country | 2. Promote Cyber security Education |
|---|---|
| USA | (1)To develop a ready Cyber Mission Force and associate cyber workforce [29].<br>(2)It will develop policies to support the National Initiative for Cyber security Education.<br>(3)To take steps for the necessary knowledge, training and other resources to countries seeking to technical and cyber security capacity. It has arranged different kinds of programs to help other nations to gain the resources and skills [30].<br>(4)Increase the state's ability to fight cybercrime including training for law enforcement, forensic specialists etc. |
| Japan | (1)To raise awareness activities starting from the elementary and middle school education stages, and implement participatory awareness raising projects.<br>(2) Promotion of voluntary activities of private enterprises and educational organizations [31].<br>(3)To make people understand about the awareness of cyber security.<br>(4) To take actions to protect critical infrastructure, establishing an institute to evaluate and issue certificates for industrial control systems, adding more categories to critical infrastructure if cyber-attacks cause significant impact on the lives of citizens. |
| Singapore | (1)To develop a vibrant cyber security ecosystem comprising a skilled workforce, technologically-advanced campaigns and strong research collaborations so that it can support Singapore's cyber security needs and be a source of new economic growth [32].<br>(2)Scholarship programs and industry-oriented curriculums will be introduced, while up-skilling and re-skilling opportunities for mid-career professionals will be provided through initiatives such as the Cyber Security Associates and Technologists Program. |
| India | (1)To conduct several awareness and training programs on cybercrimes for law enforcement agencies including those on the use of cyber Forensics Software packages and the associated procedures.<br>(2) The CBI and many state police organizations are today geared to tackle cybercrime through specialized cybercrime cells that they have set up.<br>(3) To foster education and training programs both in formal and informal sectors to support the Nation's cyber security needs and build capacity.<br>(4)To establish cyber security concept labs for awareness and skill development in key areas.<br>(5)To promote and launch a comprehensive national awareness program on security of cyberspace. |
| Malaysia | (1)Standardize and Coordinate cyber security awareness and education programs across all element of the CNII.<br>(2)Establish effective mechanism for cyber security knowledge dissemination at the national level.<br>(3)Develop foster and maintain a national culture of security.<br>(4)Develop a standard business continuity management framework.<br>(5)Implement an evaluation/certification program for cyber security products and systems. |
| Bangladesh | (1)To increase the capability of cyber security professionals in Managerial, Technical and Information Assurance areas.<br>(2)To add cyber security awareness to the national education curriculum as a way of spreading knowledge to pupils and their relatives.<br>(3)Deliver or manage commercial delivery of training or certification examinations.<br>(4)Invest in mainstream cyber security education and research.<br>(5)It may be helpful to train senior policymaker , governmental officials about the threats to electronic networks (for example, how the national banking system could be attacked) and about the threats posed by electronic networks (for example, the use of the internet to locate vulnerable children for sexual trafficking)<br>(6) To raise awareness about cyber threats, a national program should exist [28]. |

Table. 3. Comparison in Ensuring on going Risk Assessment

| Country | 3. Ensuring Ongoing Risk Assessment |
|---------|--------------------------------------|
| USA | (1)To identify threats before they can impact U.S. national security. The Defence Department continues to spread out and accomplish these solutions through continuous network monitoring, improved cyber security training for the workforce, and improved methods for identifying, reporting, and tracking suspicious behaviour [34].<br>(2)To watch, warning and incident response through exchanging information with trusted networks of international partners, The US Government actively participates in it.<br>(3) Implementing policies and protocols, it will take up to create a culture of awareness to anticipate, detect, and respond to insider threats before they have an impact. |
| Japan | (1)To establish a mechanism to implement a risk-based approach.<br>(2)It needs to continue the measures being carried out by each individual actor, while implementing handling with appropriate and timely allocation of resources for responding to ever-changing risks. |
| Singapore | (1) Through identification and prioritization of cyber risks and CIIs through risk assessments, vulnerability assessments and system reviews.<br>(2) Implement across all critical sectors, a CII Protection Program with robust and systematic cyber risk management processes. A key part of the CII Protection Program is to grow a culture of cyber risk awareness across all levels of a CII organization.<br>(3) Continuous measurement of performance through process audits and cyber security exercises [35]. |
| India | (1)To qualify implementation of global security best practices in formal risk assessment and risk management processes, business continuity management and cyber crisis management which is planned to reduce the risk of disruption and improve the security posture.<br>(2)To create awareness of the threats, vulnerabilities and consequences of bitch of security among entities for managing supply chain risks related to IT (products, systems or services) procurement. |
| Malaysia | (1)Assesses and Identifies cyber security threats exploiting vulnerabilities and risks across the CNII.<br>(2)Establish formal and encourage informal information sharing exchanges [36]. |
| Bangladesh | (1)To strengthen resilience of Critical Information Infrastructures (CII) [33].<br>(2)To focus on tackling threats most likely to prevent government agencies and businesses from carrying out critical missions.<br>(3) To provide wider participation in analysis, warning, information gathering, vulnerability reduction, mitigation and recovery. Inevitably, the government needs to work with the private sector to coordinate a national response.<br>(4) To develop a method to share information about cyber-attacks, threats and vulnerabilities with all over the people of the world.<br>(5) Providing national major incident response capacity in an event of significant attacks on critical infrastructure. |

Table. 4. Comparison in Promoting Counter Cybercrime Policy

| Country | 4. Promote Counter Cybercrime Policy |
|---------|--------------------------------------|
| USA | (1) The Defence Department will draw on best-practices to counter the proliferation of destructive malware within the international system to work with the Department of State and other agencies of the U.S. government as well as U.S. allies and partners [38].<br>(2)DoD will work in collaboration with the intelligence community to develop the data schema, databases, algorithms, and modelling and simulation (M&S) capabilities necessary to assess the effectiveness of cyber operations<br>(3) To expand companies' participation in threat information sharing programs, such as the Cyber Security/Information Assurance program.<br>(4)Protect intellectual property including commercial trade secrets from theft and industrial espionage. |
| Japan | (1)The Japanese National Cyber-Forensics and Training Alliance (NCFTA) will take measures for sharing information through cooperation with the private sector including the "Council to prevent Unauthorized Communications" as a cyber-intelligence measure.<br>(2)Japan has ratified the convention on Cybercrime and will work to strengthen rapid and effective mutual investigations and other cooperation between law enforcement agencies [40]. |
| Singapore | (1)Prevention is therefore still the key strategy to counter the threat of cybercrime. The NCAP will prioritize educating and empowering the public to be safe in cyberspace.<br>(2)To put together businesses and the community to make cyberspace safer, by countering cyber threats, combating cybercrime and protecting personal data. |
| India | Having reactive and piecemeal, India's response to cyber threats so far. India has relied either on the information of a new agency or a coordinator committee after every major cyber-attack or intelligence failure. Complementing these actions, India's department of Electronics and Information Technology (DEITY), under the aegis of Ministry of Communication and Information Technology (MCIT), released the country's maiden National Cyber Security Policy (NCSP) on 02 Jul 2013 [39]. |
| Malaysia | -- |
| Bangladesh | (1)To mobilize business and the community to make cyber space safer by countering cyber threats, combating cybercrime and protecting personal data [37].<br>(2) To create a national CIRT, defines the legal basis, for example, the Act defines the powers to shut down a critical infrastructure if at risk of a cyber-attack.<br>(3) Enhance law enforcement capabilities in the investigation, prevention and prosecution of cybercrimes. |

Table. 5. Comparison in Promoting Cyber Security in International Law

| Country | 5. Promote Cyber Security in International Law |
|---------|-----------------------------------------------|
| USA | (1)DoD will strengthen its international alliances and partnerships to develop combined capabilities to achieve cyber effects in support of combatant command plans [42].<br>(2) It will work with capable international partners to plan and train for cyber operations. |
| Japan | To secure stability of the use of cyberspace, it will promote international cooperation and establish framework of cooperation for Japan that will actively take part in international rulemaking. |
| Singapore | -- |
| India | -- |
| Malaysia | -- |
| Bangladesh | Additionally, to formulate Bangladesh's international cyber security positions, the |
| | Cyber security Coordinator or an equally empowered party will have to work with government departments and agencies, the private sector and academia [41]. |

Table. 6. Comparison in Forms of Regulation and Institutional Aspects

| Country | 6. Forms of Regulation and Institutional Aspects |
|---|---|
| USA | (1) To enhance cyber security through regulation and collaborative efforts between government and the private-sector, it will encourage voluntary improvements to cyber security. <br> (2) To demanding companies to improve cyber security, Congress is also considering bills that criminalize cyber-attacks to requiring companies to develop cyber security |
| Japan | (1)For the diverse entities such as the government, public, academic, industrial and private sectors in Japan, it becomes necessary for each entity to carry out its own information security measures in an independent and proactive fashion as part of its social responsibilities. <br> (2)Japan has worked towards constructing a safe and reliable cyberspace in which free flow of information is ensured by ensuring opens and interoperability of cyberspace without excessively administering or regulating it. <br> (3)To promote strengthen the basic functions of the nation related to cyberspace. <br> (4)Cyberspace-related operators will create a market through development of advanced technologies and products, cultivation of human resources with high ability and the use and application of these resources for information security measures in order to strengthen the international competitiveness of Japan's cyber security industry. |
| Singapore | (1)Aside from public education and outreach, a key method of cybercrime prohibition is to increase the difficulty of committing such offences by plugging potential loopholes in digital platforms and processes. MHA will regularly review regulatory frameworks, to ensure that cybercriminals are not able to exploit vulnerabilities in technology. <br> (2) Strengthening legislation and the criminal justice framework. |
| India | (1)To injunction periodic audit and evaluation of the adequacy and effectiveness of security of information infrastructure as may be appropriate with respect to regulatory framework. <br> (2)To enable, educate and facilitate awareness of the regulatory framework [43]. |
| Malaysia | (1)Ensure that all applicable local legislation is complementary to and harmony with international laws, threats and conventions. <br> (2) To make recommendations of the type of amendments required. This would also include addressing methods and processes of reconciling and harmonizing the legislation where general comments will be made of the current legislation. |
| Bangladesh | (1)The alignment of cybercrime legislation with the ITU Toolkit for cybercrime helps international cooperation and addresses jurisdictional and evidentiary issues. <br> (2)The cybercrime law should similarly be evaluated by the local private sector by any local affiliate of the international private sector by local non-governmental organizations, by academics, by unaffiliated interested citizens, by willing foreign governments and anyone else with a recognized interest. <br> (3)It is recommended that the text of National Cybercrime law be drafted to comply with the provisions of the convention on cybercrime (2001). |

Table. 7. Comparison in Balancing Cyber Security with Civil Liberties

| Country | 7. Balancing Cyber Security with Civil Liberties |
|---|---|
| USA | (1)DoD will develop a framework and exercise its Defence Support of Civil Authorities (DSCA) capabilities in support of DHS and other agencies and with state and local authorities to help defend the federal government and the private sector in an emergency if directed [38]. <br> (2)Preserve, enhance and increase access to an open, global Internet is a clear policy priority. <br> (3)To encourage international cooperation for effective commercial data privacy protections. <br> (4) To foster viable career paths for all military personnel performing and supporting cyber operations. |
| Japan | (1)It is important to multilaterally build and strengthen partnerships with other nations and regions that share the same basic values including the basic policy, democracy, respect for basic human rights and the rule of law. For this reason, it is necessary to carry out diplomacy that promotes a balanced approach to constructing a safe and reliable cyberspace. <br> (2)Cyberspace has provided us with a variety of positive benefits including innovation, economic growth and solutions for social issues while still ensuring freedom of expression and protection of privacy. |
| Singapore | -- |
| India | (1)To encourage use of open standards to facilitate interoperability and data exchange among different products or services. <br> (2) To enhance the availability of tested and certified It products based on open standards, it will promote a consortium of Government and Private sector. <br> (3)To facilitate identification, prioritization, assessment, remediation and protection of critical infrastructure and key resources based on the plan for protection. |
| Malaysia | -- |
| Bangladesh | (1)Fosters innovation in cyber security help to develop long-term solutions. <br> (2)Define and enforce a robust government Authentication Framework. |

Table. 8.1. Comparison in public private platform

| Country | 8. Types of Cooperation <br> *Public Private Platform* |
|---|---|
| USA | Build partnerships to defend the nation. DoD will have a framework in place to cooperate with other government agencies to conduct defend the nation operations. DoD will work with FBI, CIA, DHS and other agencies to build relationships and integrate capabilities to provide the President with the widest range of options available to respond to a cyber-attack of significant consequence to the United States. |
| Japan | Cooperation with the United States, in which Japan in an alliance based on the Japan-U.S. Security Arrangements, is vital. |
| Singapore | The Government has collaborated with the private sector to jointly develop capabilities to respond to the latest cyber threats. For instance, SPF has partnered local research institutes to develop new cybercrime investigations and forensics capabilities. |
| India | (1)Engages with various developed and developing countries, multilateral organizations for knowledge sharing, market access & diversification. <br> (2)Executes Projects to showcase strength and as premier IT-ITES Hub globally. <br> (3)Capacity building, HRD and sharing of expertise in areas like e-governance, Internet governance etc. [45]. |
| Malaysia | -- |
| Bangladesh | To facilitate sharing of cyber security assets across borders or with other nation states, Bangladesh has official recognized partnerships with some organizations like ITU, APCERT, OIC-CERT [44]. |

Table. 8.2. Comparison in Types of Cooperation inter-governmental cooperation

| Country | 8. Types of Cooperation Inter-Governmental Cooperation |
|---|---|
| USA | (1)By enhancing timeliness of information flow between DHS and critical infrastructure companies, it will improve existing public-private partnerships. |
| | (2)To improve the process to lenient security clearance processes for applicable public and private sector entities to enable the federal government to share this information at the appropriate sensitive and classified levels |
| Japan | The multi-stakeholders in cyberspace need to fulfil each of the responsibilities corresponding to their respective roles in the society while mutually cooperating and assisting with each other including international cooperation and cooperation between the public private sectors. |
| Singapore | (1)Increasing cybercrime awareness in the private sector. (2)SPF regularly engages key private sector stakeholders, such as those from the Information and Communication Technology and banking industries to enhance cybercrime prevention efforts, raises awareness of cybercrimes and encourages the adoption of good cyber hygiene practices. |
| India | (1)To facilitate and cooperation among stakeholder entities including private sector for actions related to cyber threats, vulnerabilities, breaches and adoption of best practices. (2)To create a think tank for cyber security policy inputs, discussion and deliberations [46]. |
| Malaysia | Develop the National Cyber Crisis Management framework that outlines the strategy for cyber-attacks mitigation and response among Malaysia's Critical National Information Infrastructure (CNII) through public and private collaboration and coordination. |
| Bangladesh | (1)Enable collaborative work and sharing of training courses that could help alleviate the severe shortage of skilled cyber security professionals. (2)Enable real time exchange of information about cyber threats and vulnerabilities. (3) To share knowledge with the law enforcing agencies, industry and academia, BDCERT organizes events. |

Table. 8.3. Comparison in Regional Cooperation

| Country | 8. Types of Cooperation Regional Cooperation |
|---|---|
| USA | Build new strategic partnerships in the Asia-Pacific region. The Defence Department will work with key states across the Asia-Pacific to build cyber capacity and minimize risk to U.S. and allied interests, in a manner consistent with DoD's International Cyberspace Security Cooperation Guidance [47]. |
| Japan | The country will actively participate in multi-country discussions and meetings including regional frameworks such as the ASEAN Regional Forum ("ARF") Asia-Pacific forum and other related committees in the United Nations. |
| Singapore | Singapore is at the forefront of working with foreign countries to enhance our operational cooperation against cybercrime. At the regional level, Singapore is the Association of Southeast Asian Nations (ASEAN) Voluntary Lead Shepherd on Cybercrime. |

| India | (1) Ensure its strategic and economic interests are addressed by ascertaining implication of security aspects in bilateral and multilateral trade dialogues.<br>(2)Work with 'like-minded' nations and global institutions to develop cyber norms and acceptable behaviour for operating in cyberspace. |
|---|---|
| Malaysia | -- |
| Bangladesh | -- |

Table. 8.4. Comparison in Intra-Governmental Cooperation

| Country | 8. Types of Cooperation<br>*Intra-Governmental Cooperation* |
|---|---|
| USA | (1) Work with key NATO allies to mitigate cyber risks to DoD and U.S. national interests. The Defence Department will develop these partnerships through the defence consultations that DoD holds with its key NATO allies.<br>(2) To develop and improve existing military alliances to confront potential threats in cyberspace. |
| Japan | (1)It is important to ensure that it can be used in a stable manner. In this regard, international rules need to be made for various activities which make use of cyberspace, while strengthening personnel ties in the medium and long term.<br>(2)To strengthen international collaboration in order to effectively respond to cybercrime, this can easily be carried out across national borders. |
| Singapore | (1)Step up efforts to forge strong international partnerships, given that cyber threats do not respect sovereign boundaries.<br>(2)Strong international partnerships enable countries to deal with cybercrime more effectively. Fostering regional and global cooperation, partner INTERPOL and other countries in capacity building initiatives, and bring global experts and thought leaders together to discuss the latest threats, trends and solutions in the cyber domain, and share best practices and solutions. |
| India | (1)To develop bilateral and multi-lateral relationships in the area of cyber security with other countries.<br>(2)To promote National and global cooperation among some representations like security agencies, CERTs, defence agencies and forces, Law Enforcement Agencies and the judicial systems. |
| Malaysia | (1)Encourage active participation in all relevant international cyber security bodies, panels and multinational agencies.<br>(2)Strengthen relationship amongst CERT/CSIRT in the OIC countries.<br>(3)Promote active participation in all relevant international cyber security by hosting an annual international cyber security conference. |
| Bangladesh | (1)Bangladesh has access to relevant cyber security services by ITU-IMPACT.<br>(2) To participate in international cooperation dialogue and coordination activities, it will take some steps to focus on cyber security such mutual assistance.<br>(3)Stepping up efforts to forge strong international partnerships. |

To make the Comparison understand easily, the cross section comparison of cyber security strategy in each of the eight specific categories for the mentioned countries are presented in a single table below. It is shown by a tic mark on the basis of three status criteria, whether the strategy is present, partially present (P.present), and absent.

Table. 9. Overall comparison in respect to eight comparison criteria

| Comparison criteria | | Status criteria | USA | Japan | Singa pore | India | Malay sia | Bangla desh |
|---|---|---|---|---|---|---|---|---|
| *1. Promote Cyber Security R & D* | | Present | √ | √ | √ | √ | | |
| | | P.present | | | | | √ | √ |
| | | Absent | | | | | | |
| *2. Promote Cyber security Education* | | Present | √ | √ | √ | √ | √ | √ |
| | | P.present | | | | | | |
| | | Absent | | | | | | |
| *3. Ensuring Ongoing Risk Assessment* | | Present | √ | | √ | | | √ |
| | | P.present | | √ | | √ | √ | |
| | | Absent | | | | | | |
| *4. Promote Counter Cybercrime Policy* | | Present | √ | | | | | |
| | | P.present | | √ | √ | √ | | √ |
| | | Absent | | | | | √ | |
| *5. Promote Cyber Security in International Law* | | Present | √ | | | | | |
| | | P.present | | √ | | | | √ |
| | | Absent | | | √ | √ | √ | |
| *6. Forms of Regulation and Institutional Aspects* | | Present | √ | √ | √ | √ | √ | √ |
| | | P.present | | | | | | |
| | | Absent | | | | | | |
| *7. Balancing Cyber Security with Civil Liberties* | | Present | √ | √ | √ | | | |
| | | P.present | | | | | | √ |
| | | Absent | | | | √ | √ | |
| *8. Types of Coope- ration* | *i. Public Private Platform* | Present | √ | | √ | | | √ |
| | | P.present | | √ | | √ | √ | |
| | | Absent | | | | | | |
| | *ii. Inter- Governmental Cooperation* | Present | | | | √ | | |
| | | P.present | √ | √ | √ | | | √ |
| | | Absent | | | | | √ | |
| | *iii. Regional Cooperation* | Present | | | | | | |
| | | P.present | √ | √ | √ | √ | | |
| | | Absent | | | | | √ | √ |
| | *iv. Intra- Governmental Cooperation* | Present | √ | √ | √ | √ | √ | √ |
| | | P.present | | | | | | |
| | | Absent | | | | | | |

From the eleven (11) categories and sub-categories, for a specific country, maximum number of 'Present', 'Partially Present' and 'Absent' will be 11 at best, and the summation of three options will also be 11. The chart below shows the status of 11 categories and subcategories of the strategic criteria of six countries for the option 'Present'. That is, this chart represents the total number of Strategy Criteria present in the Strategy of specific country by comparison with others.

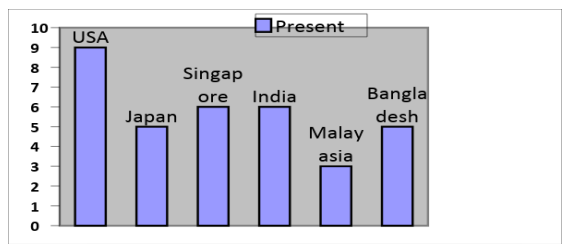

Fig 1. Chart for option 'Present' for six countries for 11 criteria

The chart below shows the status of 11 categories and subcategories of the strategic criteria of six countries for the option 'Partially Present'. That is, this chart represents the total number of Strategy Criteria partially present in the Strategy of specific country by comparison with others.

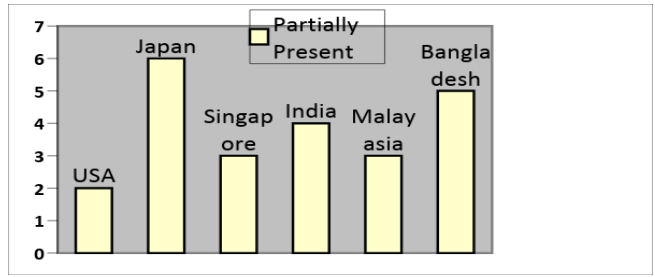

Fig 2. Chart for option 'Partially Present' for six countries for 11 criteria

The chart below shows the status of 11 categories and subcategories of the strategic criteria of six countries for the option 'Absent'. That is, this chart represents the total number of Strategy Criteria absent in the Strategy of specific country by comparison with others.

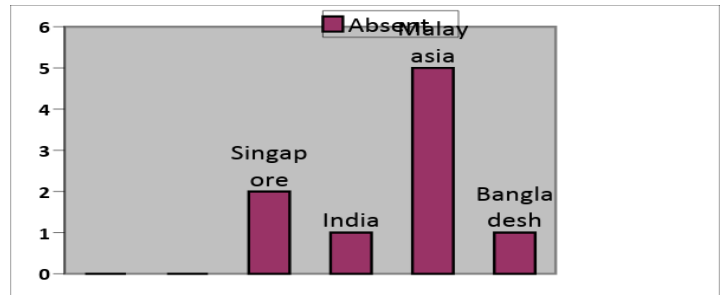

Fig 3. Chart for option 'Absent' for six countries for 11 criteria

## 6. RESULT DISCUSSION AND RECOMMENDATIONS

By analysing the comparison data from the above tables, Improvements are suggested in the following categories.

### 6.1 Research and Development

In case of research and development, Bangladesh has taken a more generalized approach than a specified one. Research and development on cyber security is listed but in which approach or what kind of initiative would be taken, have not been mentioned. That clearly explains that much thought was not given on these particular criteria when this strategy was proposed. To do further improvement in the R&D, following initiatives can be considered –R&D practice in both public and private sectors (Indian strategy), practical testing of technology (Japan Strategy) research both in technology and human science (Singapore Strategy), establishing security R&D consortium (Singapore Strategy).

### 6.2 Promote Cyber Security Education

Bangladesh has included cyber security education plan in their strategy. As it's a vision for 2021, hopefully they can be seen in implementation in upcoming years. The plan on promoting cyber security education involves training of security professionals, spreading awareness through education curriculum, certifications and training of policy makers, government officials and anyone related to the cyber security force. To further improve the security following initiatives can be considered –spreading awareness from elementary schools through including cyber security in their education (Japan), training programs both in formal and informal sectors in cyber security (India), voluntary activities of private enterprises and educational institutions to spread awareness (Japan), scholarship programs and industry oriented curriculum to encourage talents to step into cyber security (Singapore). Though the strategy of UK was not compared however, to improve educational involvement to create higher and degree-level apprenticeships to address skills gaps in essential areas, supporting the accreditation of teacher professional development in cyber security to support others learning and to understand cyber security education and provide a method of externally accrediting such individuals and to develop a Defensive Cyber Academy for cyber training can be suggested (UK 2016).

### 6.2.1 Ensure Ongoing Risk Assessment

The cyber security policy of Bangladesh shows quite promising aspects about risk assessment. In fact the policy is quite strong in comparison with other countries that are being compared with. To further improve the strategy the followings can be considered -identification and prioritization of risks and reviewing them, and continuous measurement of performance through process audits and cyber security exercises (Singapore), establish formal and encourage informal information sharing exchanges (Malaysia), engage international participation in cyber security exercises and protect intellectual property (US 2014). Use of existing powers for online offences and spread the use of Cyber-Specials to help the police (UK 2010) can also be suggested.

### 6.3 Promote Counter Cybercrime Policy

To strengthen the cyber security, Bangladesh has prepared their plan for fighting against attacks in the policy. Establishing National CIRT, creating cyber forensic division, mobilizing threats, enhancing law enforcement capabilities are some notable positive traits of the cyber security strategy of Bangladesh. To further improve its strength following can be considered -Rapid and effective mutual investigation and other cooperation between security enforcement agencies (Japan), install products on government networks that will provide assurance that software is running correctly and not being maliciously interfered with, promote an Internet Protocol (IP) reputation service to protect government digital services (this would allow online services to get information about an IP address connecting to them, helping the service make more informed risk management decisions in real time) and foster hardware and software providers to sell products with security settings activated as default, requiring the user to actively disable these settings to make them insecure (UK 2016).

### 6.4 Promote Cyber Security in International Law

This is another generalized section of the cyber policy of Bangladesh. To participate in the major discussions in international aspect, some empowered party will be responsible for keeping the balance between international and national interests. To promote cyber security in international law, as Bangladesh is a member country of IMPACT, there are possibilities to make additional improvements in this matter. The following steps adopted by strategically strong countries can be considered such as establish framework of cooperation and exchange information with regional or international forums such as ASEAN (Japan) and promote openness to innovation (UK 2010).

### 6.5 Forms of Regulation

The strategy suggests cybercrime legislation to be aligned with the ITU toolkit along with the public-private evaluation. To further improve the strength, educating and facilitating awareness of the regulatory framework (India) can be suggested.

## 6.6 Balancing Cyber Security with Civil Libraries

Cyber security and civil liberty should be in parallel in order to keep a non-suffocating cyber space. The cyber policy of Bangladesh is not very bright in this matter. To further improve, ensure freedom of expression and protection of privacy (Japan), support civil society to achieve reliable, secure and safe platforms (US 2014) can be suggested.

## 6.7 Types of Cooperation

Bangladesh is only behind in regional cooperation. To improve it further following can be considered, actively participating in multi country discussions and meetings of regional forums and making pact with them such as the ASEAN (Japan, Singapore), engaging key private sector stakeholders, to enhance cybercrime prevention efforts, raise awareness of cybercrimes and encourage the adoption of good cyber hygiene practice (Singapore), fosters international and regional organizations to support capacity building (UK-2010) can be suggested. Strengthen and embed a common understanding of responsible state behaviour, building on agreement that continue to promote the agreement and ensure fewer places exist where cyber criminals can act (UK 2016) can also be suggested.

## 7. CONCLUSION AND FUTURE WORK

By observing the cyber security strategy along with the other security strategies of different countries, it is clear that some of the criteria of the strategy of Bangladesh are in very good condition and some are yet to be given some more thoughts as there is scope for improvements. The strategy is somewhat strong in promoting cyber security education, ongoing risk assessments, counter cybercrime policy, forms and regulation and types of cooperation though most of these have the space for some minor improvements. However, some major improvements are required in the Research and Development, Coordination with the International law, Balancing Cyber Security with Civil Liberties and regional cooperation.

Though the study found that the cyber security strategy of Bangladesh is in a promising state, by improving the mentioned sectors and implementing the existing ones, only then this can be called as a resilient strategy that is prepared to handle most situations. However, strategy is merely a document if not utilized and implemented properly. Only the proper implementation of these can keep the country safe from a hostile cyberspace and turn that into a safer one. Although this study seeks to assess the NCSS of Bangladesh and provide suggestions to improve it, the findings and suggestions should be considered tentative. These suggestions were collected from the existing strategies of different countries. It was not justified if they are really compatible with the national interest of Bangladesh. Therefore the reliability and compatibility of these suggestions becomes the subject of some future research.

**AUTHORS**

Kaushik Sarker received his B.Sc. in Electronics and Telecommunication Engineering under Electrical Engineering and Computer Science and M.Sc. in Computer Systems and Network Engineering under Computing and Information Systems. Currently working as an Assistant Professor & Associate Head in the Department of Software Engineering. His research interests include Digital Image Processing and Computer Vision, Robotics, Embedded Artificial Intelligence Systems, Data Science and Cyber Security.

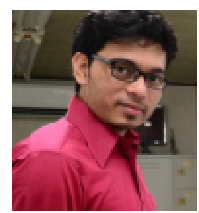

**Hasibur Rahman** is a software Engineer who specializes in SAP technologies, currently working for a renowned Pharmaceutical industry in Dhaka. He received his BSc from Daffodil International University in software engineering. He has a very keen interest on cyber security and related fields. The same interest led his work in this research which is a part of his final year thesis.

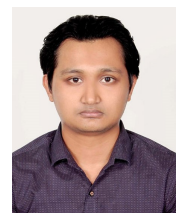

**Khandaker Farzana Rahman** was born in 10[th] December, 1993. She has completed her Bsc in Software Engineering from Daffodil International University. Currently she is working a Software Engineer in a reputed IT company at Dhaka Bangladesh. This publication is the part of her thesis during her bachelor degree study.

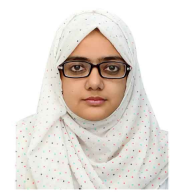

Md. Shohel Arman is a lecturer and Alumni of Department of Software Engineering under Faculty of Science & Information Technology in Daffodil International University, Dhaka, Bangladesh. He is an energetic and focused man since his student life. Currently he is focusing on his interested research area. His research interests are Machine learning, data mining, Internet of things (IOT), Cyber security and management information system (MIS).

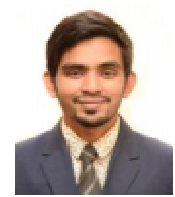

**Saikat Biswas** was born in Bangladesh, in 1994. Now, he is pursuing his under graduate degree from the Department of Software Engineering at Daffodil International University (DIU). Currently he is working as a researcher in Cyber security Centre, DIU. Research interests include Information Security, Cyber Security.

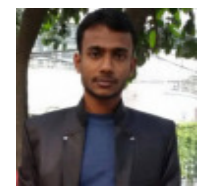

Prof. Dr. Touhid Bhuiyan is the Director of Cyber Security Centre, Daffodil International University (DIU), Bangladesh. He is also the Head and Professor of Software Engineering department of the same University. His research interests are in cyber security, intelligent recommendations, social network, trust management, big data analytics, e-Learning etc. He is the recipient of Australian Postgraduate Award (APA) and Deputy Vice-Chancellor's Initiative Scholarship from QUT, Australia.

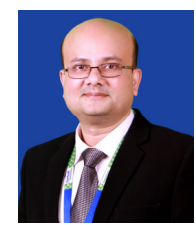